\def\nn{\nonumber}
\newcommand{\vsig}{\mbox {\boldmath $\sigma$\unboldmath}}
\newcommand{\vp}{\mbox {\boldmath $p$\unboldmath}}
\newcommand{\vq}{\mbox {\boldmath $q$\unboldmath}}
\newcommand{\vr}{\mbox {\boldmath $r$\unboldmath}}
\documentclass{svjour3}                    
\smartqed 
\usepackage{graphicx}
\begin{document}

\title{Configuration Mixing of Quark States in Baryons\footnote{Manuscript dedicated to Professor
Willibald Plessas for his sixtieth birthday. \\
Published in: Few-Body Syst {\bf 47}, 105 (2010).
}}

\author{B. Saghai Saghai \\Institut de Recherche sur les lois Fondamentales de l'Univers,\\ 
              DSM/Irfu, CEA/Saclay, 91191 Gif-sur-Yvette, France\\
              \email{bijan.saghai@cea.fr}\\ \\
and \\ \\
Z. Li \\
SGT Inc, 7701 Greenbelt Road, Greenbelt, MD. 20770, USA
}


\maketitle

\begin{abstract}
Mixing angles are used to describe the $SU(6)\otimes O(3)$ symmetry breaking 
in $\left[70,~1^-\right]$ multiplet in the sector of the lowest mass nucleon resonances, 
which are investigated extensively in constituent quark models for baryon spectroscopy. 
The transition amplitudes for the meson photoproduction off nucleon can also be expressed 
in terms of the mixing angles to take into account the configuration mixing. Those amplitudes
are derived as a function of the mixing angles between 
$|N^2P_M{J}^-\rangle$ and $|N^4P_M{J}^-\rangle$ states, with $J$=1/2 and 3/2, for the processes 
$\gamma p \to \eta p,~K^+ \Lambda,~K^+ \Sigma^\circ,~K^\circ \Sigma^+$. 
The present status of our knowledge on the mixing angles between $S_{11}(1535)$ and $S_{11}(1650)$ ($\theta _{S}$), as well as between $D_{13}(1520)$ and $D_{13} (1700)$ ($\theta _{D}$) 
is reported. 
Since these resonances play very important role in the threshold region for both $\eta$ 
and kaon production mechanisms, they are expected to provide crucial tests of different quark 
models for the baryon spectroscopy.
\end{abstract}
\section{Introduction}
\label{intro}
Discovery of the first baryon resonance, $\Delta$, goes back to 1952, when Fermi and collaborators
released~\cite{Anderson:1952nw} unexpected experimental results on the $\pi ^\pm p$ interactions, 
followed by Impulse approximation~\cite{Brueckner:1952zz}
and phase shift analysis~\cite{Anderson:1953,Anderson:1955zz,Ashkin:1956}. 
In the late 1950s and early 1960s the road to {\it The Eightfold way} was paved and by the mid-sixties a proliferation
of discovery of mesons and baryons, including nucleon resonances, received a genuine classification,
based on the SU(3) symmetry and the concept of quarks~\cite{GellMann:1964xy,Zweig:1964jf}, 
as the elementary blocks of hadrons.

A natural extension of the SU(3) flavor symmetry, taking into account the fermion nature of quarks, lead then
to the product subgroup $SU(3)\otimes SU(2)$, and hence to an $SU(6)$ flavor-spin symmetry structure of the
strongly interacting particles~\cite{SU6}. Finally, the intrinsic spin group $SU(2)$ and the internal symmetry 
group $SU(3)$, were complemented with a group of rotations in the three-dimensional space $O(3)$ symmetry, 
suggesting the invariance of strong interactions under the $SU(6)\otimes O(3)$ symmetry~\cite{Mahanthappa:1965wm}.

The early works by Copley {\it et al.}~\cite{Copley:1972tu} and 
Feynman {\it et al.}~\cite{Feynman:1971wr} on the pion photoproduction, provided the first clear evidence of the 
underlying $SU(6)\otimes O(3)$ structure of the baryon spectrum.

Those classifications initiated dynamical approaches with regards to the baryon spectroscopy, the
outcomes of which go beyond the known resonances~\cite{Amsler:2008zzb}, predicting still undiscovered or "missing" 
resonances.

Here, we concentrate on the features of the baryon spectroscopy arising from the configuration mixings. 
For the $S$- and $D$-wave resonances classified as $\left[70,~1^-\right]$ multiplet, 
the configuration mixings can be expressed as the mixing angles, which have been predicted by various quark 
models based on the baryon mass spectrum.
 
In this paper, we present a frame to investigate the configuration mixings in meson photoproduction within a 
constituent quark model and generated by the $SU(6)\otimes O(3)$ symmetry breaking. The transition amplitudes
for the resonances belonging to $\left[70,~1^-\right]$ multiplet are expressed in terms of the mixing angles, 
which can be extracted from meson photoproduction. 
Comparing to the studies~\cite{Isgur:1977ef,Isgur:1978xb,Isgur:1979be,Koniuk:1979vy,Capstick:1993kb,Bijker:1994yr,Capstick:2000qj,Giannini:2001kb,Melde:2008yr} 
on the baryon spectroscopy and transition amplitudes that are static, meson photoproduction 
processes offer additional tests on the mixing angles. 
In particular, the close to threshold behavior of 
$\eta$~ \cite{Saghai:2001yd,Zhao:2000iz,He:2007mw,He:2008ty,He:2008uf,Chiang:2002vq,Arndt:2008zz,Anisovich:2005tf,Giessen,Shyam:2008fr}
and, to a lesser extent, of kaon photoproduction~\cite{Saclay-K,Sarantsev:2005tg,Shklyar:2005xg}
reactions are largely dominated by the $S$- and $D$-wave resonances, which are very sensitive to the 
values and signs of the mixing angles.
For the latter reaction, a "`selection rule" has recently been suggested~\cite{Zhao:2006an} 
for $N^*$ resonances in the presence of QCD mixing effects.

The paper is organized as follows. In Section~\ref{sec:1} first the expressions for configuration 
mixing and the related angles are recalled (Sec.~\ref{sec:11}), then we briefly present a chiral constituent quark model,
introduce a $SU(6)\otimes O(3)$ symmetry breaking coefficient, and relate it to the configuration mixing
angles {\it via} a constant $\cal {R}$ (Sec.~\ref{sec:photo}). The main novelty of the present work is
reported in Sec.~\ref{sec:R}, where we derive explicit expressions for the constant $\cal {R}$, establishing
relations among the $SU(6)\otimes O(3)$ symmetry breaking coefficients and the mixing angles, within pseudoscalar mesons
photoproduction processes. Section~\ref{sec:MixAng} is devoted to the determination of the mixing angles, and
conclusions are given in Sec.~\ref{sec:c}. 
%
%
\section{Configuration Mixing}
\label{sec:1}
%
For the $SU(6)\otimes O(3)$ states, we use the general notation $X(^{2S+1}L_\pi)_{J^P}$, with 
$X \equiv N$ or $\Delta$, $S$ the quark spin, $L=S,~P,~D,...$ the orbital angular momentum, 
$\pi \equiv S,~M,~A$ the permutation symmetry (symmetric, mixed, antisymmetric) of the spatial wave function, and $J^P$ the state's total angular momentum and parity. The wave functions for isospin-1/2 resonances with masses below 2 GeV are given in Table~\ref{tab:Res}.

Here, the most relevant configuration mixings are those of the
two $S_{11}$ and the two $D_{13}$ states around 1.5 to 1.7 GeV (Table~\ref{tab:Res}). 
The configuration mixings can be expressed in terms of the mixing angle
between the two $SU(6)\otimes O(3)$ states $|N(^2P_M)_{J^P}>$  and 
$|N(^4P_M)_{J^P}>$, with the total quark spin $J$=1/2 and 3/2;  
%
%
\begin{eqnarray}\label{eq:MixS}
\left(\matrix{|S_{11}(1535)> \cr
|S_{11}(1650)>\cr}\right) &=&
\left(\matrix{ \cos \theta _{S} & -\sin \theta _{S}\cr
\sin \theta _{S} & \cos \theta _{S}\cr}\right) 
%
\left(\matrix{|N(^2P_M)_{{\frac 12}^-}> \cr
|N(^4P_M)_{{\frac 12}^-}>\cr}\right),  
\end{eqnarray}
%
%
and  
%
%
\begin{eqnarray}\label{eq:MixD}
\left(\matrix{|D_{13}(1520)> \cr
|D_{13}(1700)>\cr}\right) &=&
\left(\matrix{ \cos \theta _{D} & -\sin \theta _{D}\cr
\sin \theta _{D} & \cos \theta _{D}\cr}\right) 
%
\left(\matrix{|N(^2P_M)_{{\frac 32}^-}> \cr
|N(^4P_M)_{{\frac 32}^-}>\cr}\right),  
\end{eqnarray}
%
%
where $\theta_{S}$ and $\theta_{D}$ are the mixing angles between
the $S_{11}$ resonances and the $D_{13}$ resonances, respectively.
\begin{center}
\begin{table}[ht!]
\caption{Nucleon resonances with their assignments in $SU(6)\otimes O(3)$ configurations. 
}
\label{tab:Res}       
\begin{tabular}{llllll}
\hline
\noalign{\smallskip}
State & $S_{11}(1535)$ & $S_{11}(1650)$& $D_{13}(1520)$ & $D_{13}(1700)$ & $D_{15}(1675)$ 
\\[1ex]
$SU(6)\otimes O(3)$& $N(^2P_M)_{\frac 12^-}$ &$N(^4P_M)_{\frac 12^-}$ & $N(^2P_M)_{\frac 32^-}$ & 
$N(^4P_M)_{\frac 32^-}$ & $N(^4P_M)_{\frac 52^-}$ 
\\ \\
State & 
$P_{13}(1720)$ & $F_{15}(1680)$ & $P_{11}(1440)$ & $P_{11}(1710)$ 
\\ [1ex]
$SU(6)\otimes O(3)$& $N(^2D_S)_{\frac 32^+}$ & $N(^2D_S)_{\frac 52^+}$ & $N(^2S^\prime_S)_{\frac 12^+}$ &
$N(^2S_M)_{\frac 12^+}$  
\\ \\
State & $P_{13}(1900)$ & $F_{15}(2000)$     
\\ [1ex]  
$SU(6)\otimes O(3)$& $N(^2D_M)_{\frac 32^+}$ & $N(^2D_M)_{\frac 52^+}$
\\
\noalign{\smallskip}\hline
\end{tabular} 
\end{table} 
\end{center}
%

%
\subsection{Chiral constituent quark approach}
\label{sec:11}
In this Section, we investigate the manifestations of the $SU(6)\otimes O(3)$ symmetry breaking
in the $\eta$ and kaon photoproduction processes, where contributions from $S$- and $D$-wave
resonances are very important in the threshold region.
First we recall the main content of the theoretical approach used here, a chiral constituent quark model~\cite{Li:1997gd}, based on an effective chiral Lagrangian~\cite{Manohar:1983md},
\begin{eqnarray} \label{eq:lg}
\mathcal{L}=\bar{\psi}[\gamma_{\mu}(i\partial^{\mu}+V^{\mu}+\gamma_5A^{\mu})-m]\psi
+\cdot\cdot\cdot,
\end{eqnarray}
where vector ($V^{\mu}$) and axial ($A^{\mu}$) currents read,
\begin{eqnarray}
V^{\mu} =
 \frac{1}{2}(\xi\partial^{\mu}\xi^{\dag}+\xi^{\dag}\partial^{\mu}\xi)~,~
%
~
A^{\mu}=
 \frac{1}{2i}(\xi\partial^{\mu}\xi^{\dag}-\xi^{\dag}\partial^{\mu}\xi),
\end{eqnarray}
with $ \xi=\exp{(i \Pi /f_m)}$ and $f_m$ the meson decay constant.
$\psi$ and $\phi_m$ are the quark and meson fields, respectively
The field $\Pi$ is a $3\otimes 3$ matrix,
\begin{equation}\label{eq:Pi}
\Pi=\left(\matrix{
\frac 1{\sqrt {2}} \pi^\circ+\frac 1{\sqrt{6}}\eta & \pi^+ & K^+ \cr 
\pi^- & -\frac 1{\sqrt {2}}\pi^\circ+\frac 1{\sqrt {6}}\eta & K^\circ \cr 
K^- & \bar {K}^\circ &-\sqrt{\frac 23}\eta \cr}
\right),
\end{equation}
in which the pseudoscalar mesons, $\pi$, $\eta$, and $K$, are treated
as Goldstone bosons so that the Lagrangian in Eq.~(\ref{eq:lg}) 
is invariant under the chiral transformation.  
Therefore, there are four components for the photoproduction of
pseudoscalar mesons based on the QCD Lagrangian,
\begin{eqnarray}\label{eq:Mfi}
{\cal M}_{fi}&=&\langle N_f| H_{m,e}|N_i \rangle + \sum_j\bigg \{
\frac {\langle N_f|H_m |N_j\rangle \langle N_j |H_{e}|N_i\rangle
}{E_i+\omega-E_j}+ \nn \\ && \frac {\langle N_f|H_{e}|N_j\rangle
\langle N_j|H_m |N_i\rangle }{E_i-\omega_m-E_j}\bigg \}+{\cal M}_T ,
\end{eqnarray}
where $N_i(N_f)$ is the initial (final) state of the nucleon, and
$\omega (\omega_{m})$ represents the energy of incoming (outgoing)
photons (mesons). 
The pseudovector and electromagnetic couplings at the tree level are given, 
respectively, by the following standard expressions:
\begin{eqnarray}
H_m~=~\sum_j \frac 1{f_m} {\bar \psi}_j\gamma_\mu^j\gamma_5^j \psi_j
\partial^{\mu}\phi_m ~,~
%
%
H_e~=~-\sum_j e_j \gamma^j_\mu A^\mu ({\bf k}, {\bf r}) ,\label{eq:He}
\end{eqnarray}
with $j$ the constituent quark index, $\mu$ the space-time coordinate index, and
 $A^\mu ({\bf k}, {\bf r})$ the electromagnetic field.
The first term in Eq.~(\ref{eq:Mfi}) is a seagull
term.
The second and third terms correspond to the {\it s-} and {\it u-}channels,
respectively. 
The last term is the {\it t-}channel contribution.

In this paper we focus on the nucleon resonance contributions. 
For {\it s-}channel, the transition amplitudes for a resonance are given by the following
expression~\cite{Saghai:2001yd,Li:1997gd}:
\begin{equation}
\label{8} {\cal M}_{N^*}=\frac
{2M_{N^*}}{s-M_{N^*}^2-iM_{N^*}\Gamma({\bf q})}e^{-\frac {{\bf
k}^2+{\bf q}^2}{6\alpha^2}} {\cal O}_{N^*},
\end{equation}
where $\sqrt {s}~\equiv~W~=~E_N+\omega_{\gamma}=E_f+\omega_m$ is the
total centre-of-mass energy of the system, $M_{N^*}$ the mass of the resonance, and ${\cal O}_{N^*}$ is determined by 
the structure of each resonance and expressed as CGLN amplitudes. The quantity ${\cal O}_{N^*}$ is also dependent
on the photoproduction processes, and explicit expressions for ${\cal O}_{N^*}$ can be found in 
Ref.~\cite{Li:1997gd} in the $SU(6)\otimes O(3)$ symmetry limit for $\pi$, $\eta$, and
kaon photoproduction. The $\Gamma({\bf q})$ in
Eq.~(\ref{8}) is the total width of the resonance, and a function
of the final state momentum ${\bf q}$.
For a resonance decaying into a two-body final state with relative angular momentum $l$,
the decay width $\Gamma({\bf q})$ is
\begin{equation}\label{40}
\Gamma({\bf q})= \Gamma_{N^*} \frac {\sqrt {s}}{M_{N^*}} \sum_{i} x_i 
\left (\frac {|{\bf q}_i|}{|{\bf q}^{N^*}_i|}\right )^{2l+1} 
\frac {D_l({\bf q}_i)}{D_l({\bf q}^{N^*}_i)},
\end{equation}
with 
\begin{equation}\label{41}
|{\bf q}^{N^*}_i|=\sqrt{\frac 
{(M_{N^*}^2-M_B^2+m_i^2)^2}{4M_{N^*}^2}-m_i^2},
\end{equation}
and 
\begin{equation}\label{42}
|{\bf q}_i|=\sqrt{\frac 
{(s-M_B^2+m_i^2)^2}{4s}-m_i^2},
\end{equation}
where $x_i$ is the branching ratio of the resonance decaying into a 
meson with mass $m_i$ and a baryon ($M_B$), and $\Gamma_{N^*}$ is the total decay width 
of the {\it s}-channel resonance with the mass $M_{N^*}$.  
The fission barrier function $D_l({\bf q})$ in Eq. (\ref{40}) is wavefunction 
dependent. Here we use
\begin{equation}\label{43}
D_l({\bf q})=e^{-\frac {{\bf q}^2}{3\alpha^2}},
\end{equation}
which is independent of $l$.

Finally, in order to introduce the $SU(6)\otimes O(3)$ symmetry breaking, 
the transition amplitude ${\cal O}_{N^*}$ is replaced by the following
substitution relation~\cite{Li:1998ni}:
\begin{equation}
\label{subO}
{\cal O}_{N^*} \to C_{N^*}{\cal O}_{N^*},
\end{equation}
where  coefficients $C_{{N^*}}$ measure the discrepancies between 
the theoretical results and the experimental data and show the extent 
to which the $SU(6)\otimes O(3)$ symmetry is broken in the photon induced processes. 
In the $SU(6)\otimes O(3)$ symmetry limit, 
$C_{N^*}$~=~0 for ${S_{11}(1650)} $, ${D_{13}(1700)}$, and 
${D_{15}(1675)} $ resonances, and $\left|C_{N^*}\right|$~=~1 for the other
resonances, in the mass range $\approx$ 1.5 - 2.0 GeV, given in Table~\ref{tab:Res}.  
In the following Section we derive expressions relating the $C_{{N^*}}$ coefficients to the
mixing angles.


\subsection{Mixing angles and pseudoscalar mesons photoproduction}
\label{sec:photo}


The scattering amplitudes ${\cal O}_{N^*}$ expressed in terms of the 
product of the photon and meson transition amplitudes are
\begin{eqnarray}\label{eq:MixAR}
{\cal O}_{N^*} ~\propto ~<N|H_m| N^*><N^*|H_e|N>,
\end{eqnarray}
with $H_m$ and $H_e$ the meson and photon transition operators,
respectively. Using the above equations, one finds
for the resonance ${S_{11}(1535)}$ 
\begin{eqnarray}\label{eq:MixAS1}
{\cal O}_{S_{11}} &\propto& 
<N|H_m \left [\cos \theta _{S}
 |N(^2P_M)_{{\frac 12}^-}> - 
\sin \theta _{S}
|N(^4P_M)_{{\frac 12}^-}> \right ] 
\nonumber\\
&& 
 \left [\cos \theta _{S} <N(^2P_M)_{{\frac 12}^-}| -
\sin \theta _{S} <N(^4P_M)_{{\frac 12}^-}|\right ]  
 H_e|N>,
\end{eqnarray}

In this approach, the photon transition amplitude $<N(^4P_M)_{{\frac 12}^-}|H_e|N>$
vanishes~\cite{Li:1997gd} due to the Moorhouse selection rule~\cite{Moorhouse:1966jn},
So, Eq.~(\ref{eq:MixAS1}) becomes
\begin{eqnarray}\label{eq:MixAS2}
{\cal O}_{S_{11}}&\propto& (\cos^2 \theta _{S} - {\cal {R}}
\sin 
\theta _{S}\cos \theta _{S}) 
<N|H_m|N(^2P_M)_{{\frac 12}^-}>
\nonumber\\
&& <N(^2P_M)_{{\frac 12}^-}|H_e|N>,
\end{eqnarray}
where $<N|H_m|N(^2P_M)_{{\frac 12}^-}> <N(^2P_M)_{{\frac 12}^-}|H_e|N>$
determines~\cite{Li:1997gd} the CGLN amplitude for the 
$|N(^2P_M)_{{\frac 12}^-}> $ state, and the ratio
\begin{eqnarray}\label{eq:MixR}
{\cal {R}} =  \frac {<N|H_m|N(^4P_M)_{{\frac 12}^-}>}
{<N|H_m|N(^2P_M)_{{\frac 12}^-}>},
\end{eqnarray} 
is a constant determined by the $SU(6)\otimes O(3)$ symmetry. 

The configuration mixing coefficients can be related to the configuration mixing angles 
\begin{eqnarray}
C_{S_{11}(1535)} &=& \cos {\theta _{S}} ( \cos{\theta _{S}} - \cal {R}_{S} 
\sin{\theta _{S}}),\label{eq:MixS15} \\
C_{S_{11}(1650)} &=& \sin {\theta _{S}} (\cal {R}_{S} \cos{\theta _{S}} + 
\sin{\theta _{S}}),\label{eq:MixS16} \\
C_{D_{13}(1520)} &=& \cos \theta _{D} (\cos\theta _{D} -\cal {R}_{D}
\sin\theta _{D}),\label{eq:MixD15} \\
C_{D_{13} (1700)} &=& \sin \theta _{D} (\cal {R}_{D} \cos\theta _{D} + 
 \sin\theta _{D}).\label{eq:MixD17}
\end{eqnarray}

Using the meson transition operator $H_m$ from the Lagrangian intervening in deriving 
the CGLN amplitudes in the quark model, we proceed to the  calculation of the ${\cal {R}}$ 
constant for the $S_{11}$ and $D_{13}$ resonances.
\subsection{Calculation of the ${\cal {R}}$ constant}
\label{sec:R}
The wave functions of the negative-parity $L$=1 nucleon resonances in CQM have the following expressions:
%
\begin{eqnarray}
|N(^4P_M)_{J^-}\rangle &=&  \frac {1}{\sqrt 2}\sum_{m} \langle J, \frac {1}{2} |1,m,\frac {3}{2},
\frac {1}{2}-m \rangle  (\phi ^\lambda \psi ^\lambda + \phi ^\rho \psi ^\rho) \chi ^S , \\
%
|N(^2P_M)_{J^-}\rangle &=&  \frac {1}{2}\sum_{m} \langle J, \frac {1}{2} |1,m,\frac {1}{2},\frac {1}{2}-m \rangle 
\nn \\
 && \Big[ (\phi ^\rho \chi ^\lambda + \phi ^\lambda \chi ^\rho)\psi ^\rho +
(\phi ^\rho \chi ^\rho - \phi ^\lambda \chi ^\lambda) \psi ^\lambda)\Big] , 
\end{eqnarray}
where $\psi$, $\chi$, and $\phi$ stand for the spatial, spin, and flavor wave functions (Table~\ref{tab:WF}),
 respectively.
Here, $\lambda$ and $\rho$ denote the mixed symmetric and mixed anti-symmetric flavor states, respectively.
\begin{center}
\begin{table}[ht!]
\caption{Flavor wave function for proton and hyperons.}
\label{tab:WF}       
\begin{tabular}{ccc}
\hline\noalign{\smallskip}
State & $\rho$ & $\lambda$  \\
\noalign{\smallskip}\hline\noalign{\smallskip}
p & ${\frac{1}{\sqrt 2}}$(udu-duu) & ${\frac{1}{\sqrt 6}}$(2uud-duu-udu) \\ \\
$\Lambda$ & ${\frac{1}{2 \sqrt 3}}$(usd+sdu-sud-dsu-2dus+2uds) & ${\frac{1}{2}}$(sud+usd-sdu-dsu) \\ \\
$\Sigma^+$ & ${\frac{1}{\sqrt 2}}$(suu-usu) & ${\frac{1}{\sqrt 6}}$(suu+usu-2uus) \\ \\
$\Sigma^\circ$ & ${\frac{1}{2}}$(sud+sdu-usd-dsu) & ${\frac{1}{2 \sqrt 3}}$(sdu+sud+dsu+usd-2uds-2dus) \\
\noalign{\smallskip}\hline
\end{tabular}
\end{table}
\end{center}

Moreover,
\begin{equation} 
|N_{f}\rangle = \frac {1}{\sqrt 2} 
(\phi ^\lambda _f \chi ^\lambda + \phi ^\rho _f \chi ^\rho)\psi ^S .
\end{equation}

The transition operator can be written as
\begin{equation}
H_M \approx \sum h^i_m \vsig _i . \vp _i e^{iq . r_i}
\approx h^3_m  \vsig _3 . \vp _3 e^{iq . r_3} ,
\end{equation}
with $\vsig _i$ spin operator, $\vp _i$ incident beam momentum, $\vq$ outgoing meson momentum, and $\vr_i$ spatial coordinate.
The isospin operators for the pseudoscalar mesons of interest here have the following expressions:
\begin{equation}
h^3_{K^+} = a^+_{3}(s) a_{3}(u) ,
\end{equation}
\begin{equation}
h^3_{K^\circ} = a^+_{3}(s) a_{3}(d) ,
\end{equation}
\begin{equation}
h^3_{\eta} = \Big( (a^+_{3}(u) a_{3}(u) + a^+_{3}(d) a_{3}(d) \Big) .
\end{equation}
Here, $a^+_{3}(s)$ and $a_{3}(u)$ ($a^+_{3}(d)$) are the creation and annihilation operators
for the strange and up (down) quarks, respectively.
So, we have
\begin{eqnarray}
\langle N_f | H_k |^4 N \rangle & \approx & \frac {1}{2} \sum_{m} \langle J, \frac {1}{2} |1,m,\frac {3}{2},
\frac {1}{2}-m \rangle 
\big\langle \phi ^\lambda _f | h^3_{m} | \phi ^\lambda _N \big\rangle
\big\langle \chi ^\lambda  | \vsig_{3} | \chi ^S  \big\rangle
\bullet \nn \\ 
&&
\big\langle \psi ^S _f  | \vp_{3} e^{iq . r_3}| \psi ^\lambda _N  \big\rangle ,
\end{eqnarray}
and
\begin{eqnarray}
\langle N_f | H_k |^2 N \rangle & \approx & \frac {1}{2 \sqrt 2} \sum_{m} \langle J, \frac {1}{2} |1,m,\frac {3}{2},
\frac {1}{2}-m \rangle 
 \Big [ \big\langle \phi ^\rho _f | h^3_{m} | \phi ^\rho _N \big\rangle
 \big\langle \chi ^\rho  | \vsig_{3} | \chi ^\rho  \big\rangle - \nn \\
&&
 \langle \phi ^\lambda _f | h^3_{m} | \phi ^\lambda _N  \big\rangle
 \big\langle \chi ^\lambda  | \vsig_{3} | \chi ^\lambda  \big\rangle \Big ]
 \bullet 
 \big\langle \psi ^S _f  | \vp_{3} e^{iq . r_3}| \psi ^\lambda _N  \big\rangle .
\end{eqnarray}
The matrix elements $\langle \phi^\alpha _f |h^3 _m | \phi^\alpha _N \rangle$ are given in Table~\ref{tab:ME}.

\begin{center}
\begin{table}[ht!]
\caption{Matrix elements $\langle \phi^\alpha _f |h^3 _m | \phi^\alpha _N \rangle$ for $\eta N$ and 
kaon-hyperon systems.}
\label{tab:ME}
\begin{tabular}{ccccc}
\hline\noalign{\smallskip}
State            & $\eta N$ & $K^+ \Lambda$           &  $K^+ \Sigma^\circ$     & $K^\circ \Sigma^+$  \\
\noalign{\smallskip}\hline\noalign{\smallskip}
$\alpha$ = $\rho$    & 1      & $\sqrt {\frac{2}{3}}$ &  0                    &   0               \\ \\
$\alpha$ = $\lambda$ & 1      &   0                   & ${\frac{\sqrt 2}{3}}$ & $-{\frac{2}{3}}$  \\
\noalign{\smallskip}\hline
\end{tabular}
\end{table}
\end{center}
Notice that
\begin{eqnarray}
\langle \chi ^\rho  | \vsig_{3} | \chi ^\rho  \big\rangle - 
 \langle \chi ^\lambda  | \vsig_{3} | \chi ^\lambda  \rangle =
 -4 \langle \chi ^\lambda  | \vsig_{3} | \chi ^\lambda  \rangle .
\end{eqnarray}
The ratio can be written as
\begin{eqnarray}\label{eq:MixR2}
{\cal {R}} = f_{\cal {R}}
 \frac 
{\sum_{m} \langle J, \frac {1}{2} |1,m,\frac {3}{2},\frac {1}{2}-m \rangle 
\langle \chi ^\lambda  | \vsig_{3} | \chi ^S \rangle
\bullet 
\langle \psi ^S  | \vp_{3} e^{iq . r_3}| \psi ^\lambda \rangle 
}
{\sum_{m} \langle J, \frac {1}{2} |1,m,\frac {3}{2},\frac {1}{2}-m \rangle
\langle \chi ^\lambda  | \vsig_{3} | \chi ^\lambda \rangle
\bullet 
\langle \psi ^S  | \vp_{3} e^{iq . r_3}| \psi ^\lambda \rangle
}.
\end{eqnarray} 

Calculation of the matrix elements goes as follows.
\begin{eqnarray} 
{\sum_{m} \langle J, \frac {1}{2} |1,m,\frac {3}{2},\frac {1}{2}-m \rangle 
\langle \chi ^\lambda  | \vsig_{3} | \chi ^S \rangle
\bullet 
\langle \psi ^S  | \vp_{3} e^{iq . r_3}| \psi ^\lambda \rangle
} = 
\nn \\
%
\langle J, \frac {1}{2} |1,0,\frac {3}{2},\frac {1}{2} \rangle 
\langle \chi ^\lambda _{\frac {1}{2}} | \vsig ^0 _{3} | \chi ^S _{\frac {1}{2}}\rangle
\langle \psi ^S  | \vp ^0_{3} e^{iq . r_3}| \psi ^\lambda _{1,0} \rangle +
\nn \\
\langle J, \frac {1}{2} |1,1,\frac {3}{2}, - \frac {1}{2} \rangle 
\langle \chi ^\lambda _{\frac {1}{2}} | \vsig ^+ _{3} | \chi ^S _{- \frac {1}{2}}\rangle
\langle \psi ^S  | \vp ^-_{3} e^{iq . r_3}| \psi ^\lambda _{1,1} \rangle +
\nn \\
\langle J, \frac {1}{2} | 1,-1,\frac {3}{2},  \frac {3}{2} \rangle 
\langle \chi ^\lambda _{\frac {1}{2}} | \vsig ^- _{3} | \chi ^S _{ \frac {3}{2}}\rangle
\langle \psi ^S  | \vp ^+_{3} e^{iq . r_3}| \psi ^\lambda _{1,-1} \rangle .
\end{eqnarray} 
For J=1/2:
\begin{eqnarray} 
{\sum_{m} \langle J, \frac {1}{2} |1,m,\frac {3}{2},\frac {1}{2}-m \rangle 
\langle \chi ^\lambda  | \vsig_{3} | \chi ^S \rangle
\bullet 
\langle \psi ^S  | \vp_{3} e^{iq . r_3}| \psi ^\lambda \rangle
} = 
\nn \\
\frac {2 \sqrt 2}{3 \sqrt 3}
\langle \psi ^S  | \vp ^0_{3} e^{iq . r_3}| \psi ^\lambda _{1,0} \rangle -
\frac {1}{3 \sqrt 3}
\langle \psi ^S  | \vp ^-_{3} e^{iq . r_3}| \psi ^\lambda _{1,1} \rangle +
\frac {1}{ \sqrt 3}
\langle \psi ^S  | \vp ^+_{3} e^{iq . r_3}| \psi ^\lambda _{1,-1} \rangle = 
\nn \\
\frac {2 \sqrt 2}{3 \sqrt 3} 
\big ( \langle \psi ^S  | \vp ^0_{3} e^{iq . r_3}| \psi ^\lambda _{1,0} \rangle -
{\sqrt 2} \langle \psi ^S  | \vp ^-_{3} e^{iq . r_3}| \psi ^\lambda _{1,1} \rangle
\big ).
\end{eqnarray} 
For J=3/2:
\begin{eqnarray} 
{\sum_{m} \langle \frac {3}{2}, \frac {1}{2} |1,m,\frac {3}{2},\frac {1}{2}-m \rangle 
\langle \chi ^\lambda  | \vsig_{3} | \chi ^S \rangle
\bullet 
\langle \psi ^S  | \vp_{3} e^{iq . r_3}| \psi ^\lambda \rangle
} = 
\nn \\
\frac {2 \sqrt 2}{3 \sqrt 15} 
\big ( \langle \psi ^S  | \vp ^0_{3} e^{iq . r_3}| \psi ^\lambda _{1,0} \rangle +
\frac {1}{ \sqrt 2} \langle \psi ^S  | \vp ^-_{3} e^{iq . r_3}| \psi ^\lambda _{1,1} \rangle
\big ) .
\end{eqnarray} 
Notice that
\begin{eqnarray} 
\langle \psi ^S  | \vp ^-_{3} e^{iq . r_3}| \psi ^\lambda _{1,1} \rangle =
\langle \psi ^S  | \vp ^+_{3} e^{iq . r_3}| \psi ^\lambda _{1,-1} \rangle ,
\end{eqnarray}
where $p ^\pm = p ^x \pm i p ^y$.

Similarly,

\begin{eqnarray} 
{\sum_{m} \langle J, \frac {1}{2} |1,m,\frac {1}{2},\frac {1}{2}-m \rangle 
\langle \chi ^\lambda  | \vsig_{3} | \chi ^\lambda \rangle
\bullet 
\langle \psi ^S  | \vp_{3} e^{iq . r_3}| \psi ^\lambda \rangle
} = 
\nn \\
%
\langle J, \frac {1}{2} |1,0,\frac {1}{2},\frac {1}{2} \rangle 
\langle \chi ^\lambda _{\frac {1}{2}} | \vsig ^0 _{3} | \chi ^\lambda _{\frac {1}{2}}\rangle
\langle \psi ^S  | \vp ^0_{3} e^{iq . r_3}| \psi ^\lambda _{1,0} \rangle +
\nn \\
\langle J, \frac {1}{2} |1,1,\frac {1}{2}, - \frac {1}{2} \rangle 
\langle \chi ^\lambda _{\frac {1}{2}} | \vsig ^+ _{3} | \chi ^S _{- \frac {1}{2}}\rangle
\langle \psi ^S  | \vp ^-_{3} e^{iq . r_3}| \psi ^\lambda _{1,1} \rangle .
\end{eqnarray}
For J=1/2:
\begin{eqnarray} 
{\sum_{m} \langle J, \frac {1}{2} |1,m,\frac {1}{2},\frac {1}{2}-m \rangle 
\langle \chi ^\lambda  | \vsig_{3} | \chi ^\lambda \rangle
\bullet 
\langle \psi ^S  | \vp_{3} e^{iq . r_3}| \psi ^\lambda \rangle
} = 
\nn \\
\frac {1}{3 \sqrt 3} 
\big ( \langle \psi ^S  | \vp ^0_{3} e^{iq . r_3}| \psi ^\lambda _{1,0} \rangle -
{\sqrt 2} \langle \psi ^S  | \vp ^-_{3} e^{iq . r_3}| \psi ^\lambda _{1,1} \rangle
\big ) .
\end{eqnarray} 
For J=3/2:
\begin{eqnarray} 
{\sum_{m} \langle \frac {3}{2}, \frac {1}{2} |1,m,\frac {1}{2},\frac {1}{2}-m \rangle 
\langle \chi ^\lambda  | \vsig_{3} | \chi ^\lambda \rangle
\bullet 
\langle \psi ^S  | \vp_{3} e^{iq . r_3}| \psi ^\lambda \rangle
} = 
\nn \\
- \frac {\sqrt 2}{3 \sqrt 3} 
\big ( \langle \psi ^S  | \vp ^0_{3} e^{iq . r_3}| \psi ^\lambda _{1,0} \rangle +
\frac {1}{ \sqrt 2} \langle \psi ^S  | \vp ^-_{3} e^{iq . r_3}| \psi ^\lambda _{1,1} \rangle
\big ) .
\end{eqnarray} 

Now, the constant $\cal {R}$ can be calculated for J=1/2:
\begin{eqnarray} 
R_S = f_{\cal {R}} \frac {\frac {2 \sqrt 2}{3 \sqrt 3}}{\frac {1}{3 \sqrt 3}} = 2 \sqrt 2 f_{\cal {R}},
\end{eqnarray}  
and for J=3/2:
\begin{eqnarray} 
R_D = - f_{\cal {R}} \frac {\frac {2 \sqrt 2}{3 \sqrt 15}}{\frac {\sqrt 2}{3 \sqrt 3}} = 
-  \frac {2} {\sqrt 5} f_{\cal {R}} .
\end{eqnarray} 

Below, we give the values of $f_{\cal {R}}$ 
for pseudoscalar mesons photoproduction processes, namely,
$\gamma p \to \eta p, K^+ \Lambda, K^+ \Sigma^\circ, K^\circ \Sigma^+$:
\begin{equation}\label{eq:fR}
{f_{\cal {R}}}=\left\{\begin{array} {cl} -\frac{1}{2 \sqrt{2}} & { ~~for~~ \eta N~,} \\ \\
0 & { ~~for ~~K \Lambda}~,\\ \\
-\sqrt {2} & { ~~for ~~K \Sigma~.}
\end{array}
\right. 
\end{equation}

Accordingly, the numerical values for the constant $\cal {R}$ are given in Table~\ref{tab:RSD}. 
Notice that the values of both $\cal {R}$ constants  vanish for the $\gamma p \to K^+ \Lambda$
channel, in agreement with the $\Lambda$ selection rule discussed in Ref.~\cite{Zhao:2006an}.
%
\begin{center}
\begin{table}[hb!]
\caption{Values of the ${\cal {R}}$ constant, within the Koniuk and Isgur~\cite{Koniuk:1979vy} convention, 
for the $\eta$ and kaon photoproduction processes.}
\label{tab:RSD}       
\begin{tabular}{cccc}
\hline\noalign{\smallskip}
 & $\gamma p \to \eta p$ &  $\gamma p \to K^+ \Lambda$ & $\gamma p \to K^+ \Sigma^\circ$,~$K^\circ \Sigma^+$  \\
\noalign{\smallskip}\hline\noalign{\smallskip}
${\cal {R}}_S$ & -1 & 0 & -4  \\ \\
${\cal {R}}_D$ & $ {\frac{1}{\sqrt 10}}$ & 0 & ${\frac{4}{\sqrt 10}}$  \\
\noalign{\smallskip}\hline
\end{tabular}
\end{table}
\end{center}

Notice that in the present work, we have adopted the convention introduced by Koniuk and Isgur~\cite{Koniuk:1979vy}, 
where wave functions are in line with the SU(3) conventions of de Swart~\cite{deSwart}. 
In this frame, e.g. for the  process $\gamma p \to \eta p$, the constant ${\cal {R}}_S$ gets a negative value, 
and the relevant mixing angle for the $S-$wave, $\theta_{S}$, turns out positive. 
However, in line with the Hey {\it et al.}~\cite{Hey:1974nc} analysis, 
Isgur and Karl in their early works~\cite{Isgur:1977ef,Isgur:1978xi,Isgur:1978xj,Isgur:1978wd} 
used another convention, for which ${\cal {R}}_S$ = +1 and $\theta_{S} <$~0. 
In the literature both conventions are being used, often without explicit mention of the utilized convention. 

Our final results relating the $SU(6)\otimes O(3)$ symmetry breaking coefficients and mixing angles are
presented in Table~\ref{tab:CSD}. 
%
\begin{center}
\begin{table}[ht!]
\caption{Relations among the symmetry breaking coefficients and mixing angles, 
Eqs.~(\ref{eq:MixS15}) to (\ref{eq:MixD17}),
for the $\eta$ and kaon photoproduction processes.}
\label{tab:CSD}       
\begin{tabular}{cccc}
\hline\noalign{\smallskip}
 & $\gamma p \to \eta p$ &  $\gamma p \to K^+ \Lambda$ & $\gamma p \to K^+ \Sigma^\circ$, ~$K^\circ \Sigma^+$  \\
\noalign{\smallskip}\hline\noalign{\smallskip}
$C_{S_{11}(1535)}$ & $\cos\theta_{S}$ ($\cos\theta_{S} $ + $\sin\theta_{S})$ & $\cos^2\theta_{S}$ & 
$\cos\theta_{S}$ ($\cos\theta_{S}$ + 4 $\sin\theta_{S}$) \\ \\
$C_{S_{11}(1650)}$ & $ \sin\theta_{S}$ ( - $\cos\theta_{S} $+ $\sin\theta_{S})$ & 
$\sin^2\theta_{S}$ & 
$\sin\theta_{S}$ (-4 $\cos\theta_{S}$ + $\sin\theta_{S}$) \\ \\
$C_{D_{13}(1520)}$ & $\cos\theta_{D}$ ($\cos\theta_{D} $- ${\frac{1}{\sqrt 10}} \sin\theta_{D})$ & 
$\cos^2\theta_{D}$ & 
$\cos\theta_{D}$ ($\cos\theta_{D} $- ${\frac{4}{\sqrt 10}} \sin\theta_{D})$\\ \\
$C_{D_{13}(1700)}$ & $\sin\theta_{D}$ (${\frac{1}{\sqrt 10}}\cos\theta_{D} $+ $ \sin\theta_{D})$ & 
$\sin^2\theta_{D}$ & 
$\sin\theta_{D}$ ($\sin\theta_{D} $+ ${\frac{4}{\sqrt 10}} \cos\theta_{D})$\\
\noalign{\smallskip}\hline
\end{tabular}
\end{table}
\end{center}

To end this section, we wish to emphasize that the photoproduction reactions in Table~\ref{tab:CSD} are
being extensively studied, since about one decade, both theoretically and experimentally.
The chiral constituent quark model, briefly presented in Sec.~\ref{sec:11} has been 
used to study $\gamma p \to \eta p$~\cite{Saghai:2001yd,Zhao:2000iz,He:2007mw,He:2008ty,He:2008uf}
and $\gamma p \to K^+ \Lambda$~\cite{Saclay-K} processes. 
The mixing angles, left as adjustable parameters, have been extracted~\cite{Saghai:2001yd,He:2007mw} by fitting 
 $\gamma p \to \eta p$ data~\cite{data-eta,Old}, including polarization asymmetries. 
Those models embody all nucleon resonances given in Table~\ref{tab:Res}.
In next Section, we report on those results as well as on findings by various authors with respect to 
the mixing angles. 
\section{Determination of mixing angles}
\label{sec:MixAng}
In mid-seventies, Hey {\it et al.}~\cite{Hey:1974nc} performed a comprehensive
analysis of decay rates of baryon resonances belonging to the $\left[70,~1^-\right]$ and
$\left[56,~2^+\right]$ representations of $SU(6)\otimes O(3)$ into the ground state 
$\left[56,~0^+\right]$ baryons and the pseudoscalar mesons. 

Based on the pioneer work by De Rujula {\it et al.}~\cite{De Rujula:1975ge} a non-relativistic
constituent quark approach was developed by Isgur, Karl, and 
collaborators~\cite{Isgur:1977ef,Isgur:1978xb,Isgur:1979be,Koniuk:1979vy,Isgur:1978xi,Isgur:1978xj,Isgur:1978wd}. 
Isgur and Karl, using a harmonic oscillator confining potential with the OGE interaction 
 found excellent~\cite{Isgur:1977ef,Isgur:1978xj} agreement with the 
extracted values from experimental decay rates. In the Isgur-Karl {\it et al.} approach, 
a major assumption is that
the quark dynamics is subject to the gluon field, which provides a confining potential. However, within
chiral perturbation theory, at low energy the effective degrees of freedom are mesons, instead of gluons.

Extensive investigations have been performed~\cite{Glozman:1995fu} considering the exchange of
a pseudoscalar octet between light quarks generating the hyperfine structure. That approach has been
generalized by Glozman, Plessas, Varga, and Wagenbrunn~\cite{Glozman:1997ag,Glozman:1999vd} 
embodying the exchange of a nonent of vector mesons and a scalar meson. Within that scope, Glozman and 
Riska~\cite{Glozman:1995fu}, generalizing one-pion-exchange (OPE) mechanism,  attributed 
the spin-dependent coupling between constituent quarks to Goldstone-boson-exchange (GBE). 

In addition to the above approaches, there are other versions of CQM, according to the embodied
quark dynamics, such as algebric~\cite{Bijker}, hypercentral~\cite{Giannini:2001kb,Giannini}, and 
instanton~\cite{Bonn}. 
This latter, a powerful formalism of relativistically covariant constituent quark model, has been 
extensively developed by the Bonn group~\cite{Bonn}. 
That approach is based on the three-fermion Bethe-Salpeter equation, where the confinement is 
generated by an instantaneous string-like three-body potential.
The results of those works allow the authors to account for the spectrum of known resonances, 
predict missing ones, and put forward an explanation for those not yet observed states, due to their
vanishing couplings to the $\pi N$ or $\overline{\!K}N$ systems, as suggested in 
Refs.~\cite{Koniuk:1979vy,Capstick:1992th}. 

Another covariant model~\cite{Roberts:2007ji} based on the Dyson-Schwinger equations, relates 
the confinement to the analytical properties of QCD's Schwinger functions, and offers a reliable frame to
interpret baryon data directly in terms of current quarks and gluons. Moreover, this approach 
establishes a link between the phenomenology of dressed current quarks and Lattice 
QCD~\cite{Roberts:2007ji,Alexandrou:2009xk}.

\begin{center}
\begin{table}[ht!]
\caption{Mixing angles for the two $S_{11}$ and the two $D_{13}$ states around 1.5 to 1.7 GeV.}
\label{tab:MA}       
\begin{tabular}{lccl}
\hline\noalign{\smallskip}
Approach & $\Theta _S$ (deg)  & $\Theta _D$ (deg) &  Authors (Ref.) \\
\noalign{\smallskip}\hline\noalign{\smallskip}  
Decay rates analysis  & -31.9 & +10.4 & Hey {\it et al.}~\cite{Hey:1974nc} \\ \\
OGE (H.O.)  & -32 & +6 & Isgur-Karl~\cite{Isgur:1977ef,Isgur:1978xj};~Chizma-Karl~\cite{Chizma:2002qi}\\ 
OGE (B.M.)  & -32 & +4 & Chizma~\cite{Chizma:2004wr}\\ \\ 
OPE (H.O.)  & $\pm$ 13 & $\pm$ 8 & Glozman-Riska~\cite{Glozman:1995fu}; Isgur~\cite{Isgur:1999jv}\\
OPE (H.O.)  & +26 & -53 & Chizma-Karl~\cite{Chizma:2002qi}\\ 
OPE (B.M.)  & +29 & -47 & Chizma~\cite{Chizma:2004wr}\\ \\
RCQM  & +38$\pm$4 & +10$\div$15 & Capstick-Roberts~\cite{Capstick:2004tb}\\ \\ 
$1/N_c$ expansion & +22 & +28 & Pirjol-Schat~\cite{Pirjol:2003ye}\\ \\
CQM ($\gamma p \to \eta p$) & -27 & ~+5 & Saghai-Li~\cite{Saghai:2001yd}\\
                            & -35 & +15 & He {\it et al.}~\cite{He:2007mw}\\ 
\noalign{\smallskip}\hline
\end{tabular}       
\end{table}
\end{center}

In Table~\ref{tab:MA}, results reported in the literature for the mixing angles are summarized
and compared with their experimental values~\cite{Hey:1974nc} (first row).
The rows two to six embody results from OGE~\cite{Isgur:1977ef,Isgur:1978xj,Chizma:2002qi} 
and OPE / GBE~\cite{Glozman:1995fu,Isgur:1999jv,Chizma:2002qi} approaches built on 
harmonic-oscillator (HO) basis for the orbital wave functions or the
Bag model (BM)~\cite{Chizma:2004wr}. Results obtained~\cite{Capstick:2004tb} within a relativized 
constituent quark model (RCQM), based on HO and OGE, give (row seven) comparable values for mixing 
angles as the non-relativistic CQM, albeit within a sign difference for the $S_{11}$ resonances
due to the use of Koniuk and Isgur~\cite{Koniuk:1979vy} convention. 
Notice that the other results coming  from CQM/OGE approaches use the conventions introduced by 
Hey {\it et al.}~\cite{Hey:1974nc}.

The $1/N_c$ expansion approach has also been extensively applied to the determination of mixing angles
~\cite{Pirjol:2003ye,Nc,Pirjol:2008gd}. 
The outcomes are within the following ranges: 0$^\circ \leq \Theta _S \leq$ 35$^\circ$ and 
0$^\circ \leq\Theta _D \leq$ 45$^\circ$, with typical values given in row eight. 

In the last two rows of Table~\ref{tab:MA} we report the results from our phenemenological chiral constituent
quark model investigation the $\gamma p \to \eta p$ process, where the $SU(6)\otimes O(3)$ symmetry breaking
coefficients are left as adjustable parameters. 
A first study~\cite{Saghai:2001yd} used the data base~\cite{Old} available in 2000, limited
to total center-of-mass energies $W \le$ 2 GeV, and led to $\Theta _S$ = -27$^\circ$ and  
$\Theta _D$ = +5$^\circ$. 
Since then, much copious and accurate experimental results~\cite{data-eta}
have been released up to $W \approx$ 2.6 GeV. That data base has been investigated within
a more advanced approach~\cite{He:2007mw}, embodying all known resonances. 
The extracted values 
are $\Theta _S$ = -35$^\circ$ and  $\Theta _D$ = +15$^\circ$, and turn out to be in 
good agreement with experimental values and those calculated by Isgur and 
Karl\footnote{Notice that within Isgur-Karl convention, those angles lead to the following values for the configuration 
mixing coefficients (Eqs.~(\ref{eq:MixS15}) to (\ref{eq:MixD17}), with ${\cal {R}}_S$= 1): 
$C_{S_{11}(1535)}$ = 1.14,
$C_{S_{11}(1650)}$ = - 0.14,
$C_{D_{13}(1520)}$ = 0.85, and
$C_{D_{13}(1700)}$ = 0.15. Given that the unbroken $SU(6)\otimes O(3)$ symmetry predicts for
those coefficients
$ | C_{S_{11}(1535)} |$ = $ | C_{D_{13}(1520)}|$ = 1, and
$C_{S_{11}(1650)}$ = $C_{D_{13}(1700)}$ = 0, 
then the symmetry breaking effects come out to be around 15\%.
}.
However, more investigation are needed for kaon photoproduction processes, where contributions
from the second $S_{11}$ resonance are more significant (Table~\ref{tab:CSD}) than in $\gamma p \to \eta p$.
A systematic study in both $\eta$ and kaon photoproduction reactions will provide more accurate 
information on the mixing angles.

%
\section{Conclusions}
\label{sec:c}
In this paper, within a chiral constituent quark approach, we reported on the derivation 
of relations between the transition amplitudes in the $\eta$ and kaon photoproduction 
channels and 
the two most widely investigated mixing angles ($\Theta _S$ and $\Theta _D$) for the 
resonances $S_{11}(1535)$ and $S_{11}(1650)$;  $D_{13}(1520)$ and $D_{13}(1700)$. 
The extracted mixing angles from the photoproduction process
$\gamma p \to \eta p$ are in good agreement with those obtained from the resonances decay 
data~\cite{Hey:1974nc}.

The mixing angles from the $S$- and $D$-wave resonance offer important tools to test various 
quark models, which may provide us with insights into the underlying dynamics of the quarks 
interations. 
This program requires systematic studies both on baryon spectroscopy and on the properties of 
strong, weak and electromagnetic transitions.  
A recent work~\cite{Pirjol:2008gd} puts forward a quark Hamiltonian  embodying a mix of the OGE 
and GBE interactions. Another important development in this realm comes from recent investigations 
concluding that the SU(6) symmetry breaking effects can be attributed, partly to spin- and flavor-dependent
interactions between the quarks, and partly to loop effects, emphasizing the need for careful 
treatment of mixing mechanisms~\cite{Capstick:2007tv}.
 
 Results from other approaches were briefly discussed. We found that the 
extracted mixing angle from the $\eta$ photoproduction are consistent with the results from the 
Isgur-Karl~\cite{Isgur:1977ef,Isgur:1978xj} model. 
More investigations are needed in the kaon photoproduction, where contributions from the 
second $S_{11}$ state are significant. 
 
\begin{acknowledgements}

We wish to thank Qiang Zhao for fruitful exchanges.
\end{acknowledgements}
%


\begin{thebibliography}{99}%


\bibitem{Anderson:1952nw}
  H.~L.~Anderson, E.~Fermi, E.~A.~Long and D.~E.~Nagle,
  Phys.\ Rev.\  {\bf 85}, 936 (1952).

\bibitem{Brueckner:1952zz}
  K.~A.~Brueckner,
  Phys.\ Rev.\  {\bf 86}, 106 (1952).
  
\bibitem{Anderson:1953}
  H.~L.~Anderson, E.~Fermi, R.~Martin and D.~E.~Nagle,
  Phys.\ Rev.\  {\bf 91}, 155 (1953).

\bibitem{Anderson:1955zz}
  H.~L.~Anderson, W.~C.~Davidon and U.~E.~Kruse,
  Phys.\ Rev.\  {\bf 100}, 339 (1955).

\bibitem{Ashkin:1956}
  J.~Ashkin, J.~P.~Blaser, F.~Feiner, and ~M.~O.~Stern,
  Phys.\ Rev.\  {\bf 101}, 1149 (1956).
  
\bibitem{GellMann:1964xy}
  M.~Gell-Mann and Y.~Neemam,
  {\it The eightfold way: a review with a collection of reprints},
  W. A. Benjamin Publisher (1964).

\bibitem{Zweig:1964jf}
  G.~Zweig, 
  CERN preprint 8409/Th. 412, unpublished.

\bibitem{SU6}
F.~G\"ursey, A.~Pais, and L.~A.~Radicati,
 Phys.\ Rev.\ Lett.\  {\bf 13}, 173 (1964);
%
A.~Pais,
 Phys.\ Rev.\ Lett.\  {\bf 13}, 175 (1964);
%
F.~G\"ursey and L.~A.~Radicati,
 Phys.\ Rev.\ Lett.\  {\bf 13}, 299 (1964);
 %
T. K. Kuo and Tsu Yao, 
 Phys.\ Rev.\ Lett.\  {\bf 13}, 415 (1964);
 %
 Mirza A. Baqi B\'eg and Virendra Singh,
 Phys.\ Rev.\ Lett.\  {\bf 13}, 418 (1964);
 %
 B. Sakita,
 Phys.\ Rev.\ {\bf 136}, B1756 (1964). 

\bibitem{Mahanthappa:1965wm}
  K.~T.~Mahanthappa and G.~Sudarshan,
  Phys.\ Rev.\ Lett.\  {\bf 14}, 163 (1965).

\bibitem{Copley:1972tu}
  L.~A.~Copley, G.~Karl and E.~Obryk,
  Phys.\ Rev.\  D {\bf 4}, 2844 (1971).

\bibitem{Feynman:1971wr}
  R.~P.~Feynman, M.~Kislinger and F.~Ravndal,
  Phys.\ Rev.\  D {\bf 3}, 2706 (1971).

  
\bibitem{Amsler:2008zzb}
  C.~Amsler {\it et al.}  [Particle Data Group],
  Phys.\ Lett.\  B {\bf 667}, 1 (2008).

\bibitem{Isgur:1977ef}
  N.~Isgur and G.~Karl,
  Phys.\ Lett.\  B {\bf 72}, 109 (1977).
  
\bibitem{Isgur:1978xb}
  N.~Isgur, G.~Karl and R.~Koniuk,
  Phys.\ Rev.\ Lett.\  {\bf 41}, 1269 (1978);  
  Erratum-ibid.\  {\bf 45}, 1738 (1980).

\bibitem{Isgur:1979be}
  N.~Isgur and G.~Karl,
  Phys.\ Rev.\  D {\bf 20}, 1191 (1979).
  
\bibitem{Koniuk:1979vy}
  R.~Koniuk and N.~Isgur,
  Phys.\ Rev.\  D {\bf 21}, 1868 (1980);
  Erratum-ibid.\  D {\bf 23}, 818 (1981).

\bibitem{Capstick:1993kb}
  S.~Capstick and W.~Roberts,
  Phys.\ Rev.\  D {\bf 49}, 4570 (1994).
  
\bibitem{Bijker:1994yr}
  R.~Bijker, F.~Iachello and A.~Leviatan,
  Annals Phys.\  {\bf 236}, 69 (1994).

\bibitem{Capstick:2000qj}
  S.~Capstick and W.~Roberts,
  Prog.\ Part.\ Nucl.\ Phys.\  {\bf 45}, S241 (2000).
%
\bibitem{Giannini:2001kb}
  M.~M.~Giannini, E.~Santopinto and A.~Vassallo,
  Eur.\ Phys.\ J.\  A {\bf 12}, 447 (2001). 

\bibitem{Melde:2008yr}
  T.~Melde, W.~Plessas and B.~Sengl,
  Phys.\ Rev.\  D {\bf 77}, 114002 (2008).

\bibitem{Saghai:2001yd}
  B.~Saghai and Z.~p.~Li,
  Eur.\ Phys.\ J.\  A {\bf 11}, 217 (2001).
  
\bibitem{Zhao:2000iz}
  Q.~Zhao, B.~Saghai and Z.~p.~Li,
  J.\ Phys.\ G {\bf 28}, 1293 (2002). 
  
\bibitem{He:2007mw}
  J.~He, B.~Saghai, Z.~Li, Q.~Zhao and J.~Durand,
  Eur.\ Phys.\ J.\  A {\bf 35}, 321 (2008).
  
\bibitem{He:2008ty}
  J.~He, B.~Saghai and Z.~Li,
  Phys.\ Rev.\  C {\bf 78}, 035204 (2008).
  
\bibitem{He:2008uf}
  Jun~He and B.~Saghai,
  Phys.\ Rev.\  C {\bf 80}, 015207 (2009).

\bibitem{Chiang:2002vq}
  W.~T.~Chiang, S.~N.~Yang, L.~Tiator, M.~Vanderhaeghen and D.~Drechsel,
  Phys.\ Rev.\  C {\bf 68}, 045202 (2003).

\bibitem{Arndt:2008zz}
  R.~Arndt, W.~Briscoe, I.~Strakovsky and R.~Workman,
  Eur.\ Phys.\ J.\  A {\bf 35}, 311 (2008).
  
\bibitem{Anisovich:2005tf}
  A.~V.~Anisovich, A.~Sarantsev, O.~Bartholomy, E.~Klempt, V.~A.~Nikonov and U.~Thoma,
  Eur.\ Phys.\ J.\  A {\bf 25}, 427 (2005).

\bibitem{Giessen}  
  T.~Feuster and U.~Mosel,
  Phys.\ Rev.\  C {\bf 59}, 460 (1999);
%
  V.~Shklyar, H.~Lenske and U.~Mosel,
  Phys.\ Lett.\  B {\bf 650}, 172 (2007). 
 
  
\bibitem{Shyam:2008fr}
  R.~Shyam and O.~Scholten,
  arXiv:0808.0632 [nucl-th].
\bibitem{Saclay-K} 
  B.~Julia-Diaz, B.~Saghai, F.~Tabakin, W.~T.~Chiang, T.~S.~Lee and Z.~Li,
  Nucl.\ Phys.\  A {\bf 755}, 463 (2005);  
%
  B.~Julia-Diaz, B.~Saghai, T.~S.~Lee and F.~Tabakin,
  Phys.\ Rev.\  C {\bf 73}, 055204 (2006);
%
  B.~Saghai, J.~C.~David, B.~Julia-Diaz and T.~S.~Lee,
  Eur.\ Phys.\ J.\  A {\bf 31}, 512 (2007).
  
\bibitem{Sarantsev:2005tg}
  A.~V.~Sarantsev, V.~A.~Nikonov, A.~V.~Anisovich, E.~Klempt and U.~Thoma,
  Eur.\ Phys.\ J.\  A {\bf 25}, 441 (2005).
  
\bibitem{Shklyar:2005xg}
  V.~Shklyar, H.~Lenske and U.~Mosel,
  Phys.\ Rev.\  C {\bf 72}, 015210 (2005).
  
\bibitem{Zhao:2006an}
  Q.~Zhao and F.~E.~Close,
  Phys.\ Rev.\  D {\bf 74}, 094014 (2006). 

  
\bibitem{Li:1997gd}
  Z.~p.~Li, H.~x.~Ye and M.~h.~Lu,
  Phys.\ Rev.\  C {\bf 56}, 1099 (1997).


\bibitem{Manohar:1983md}
  A.~Manohar and H.~Georgi,
  Nucl.\ Phys.\  B {\bf 234}, 189 (1984).

\bibitem{Li:1998ni}
  Z.~p.~Li and B.~Saghai,
  Nucl.\ Phys.\  A {\bf 644}, 345 (1998).


\bibitem{Moorhouse:1966jn}
  R.~G.~Moorhouse,
  Phys.\ Rev.\ Lett.\  {\bf 16}, 772 (1966).
  
\bibitem{deSwart}
  J.~J.~de Swart,
  Rev.\ Mod.Phys.\  {\bf 35}, 916 (1963).


  
\bibitem{data-eta}
  M.~Dugger {\it et al.}  [CLAS Collaboration],
  Phys.\ Rev.\ Lett.\  {\bf 89}, 222002 (2002);
%
  V.~Crede {\it et al.}  [CB-ELSA Collaboration],
  Phys.\ Rev.\ Lett.\  {\bf 94}, 012004 (2005);
%
  T.~Nakabayashi {\it et al.},
  Phys.\ Rev.\  C {\bf 74}, 035202 (2006);
%
  O.~Bartalini {\it et al.}  [The GRAAL collaboration],
  Eur.\ Phys.\ J.\  A {\bf 33}, 169 (2007);
%
  O.~Bartholomy {\it et al.}  [CB-ELSA Collaboration],
  Eur.\ Phys.\ J.\  A {\bf 33}, 133 (2007);
%
  D.~Elsner {\it et al.}  [CBELSA Collaboration and TAPS Collaboration],
  Eur.\ Phys.\ J.\  A {\bf 33}, 147 (2007).  

\bibitem{Hey:1974nc}
  A.~J.~G.~Hey, P.~J.~Litchfield and R.~J.~Cashmore,
  Nucl.\ Phys.\  B {\bf 95}, 516 (1975).

\bibitem{De Rujula:1975ge}
  A.~De Rujula, H.~Georgi and S.~L.~Glashow,
  Phys.\ Rev.\  D {\bf 12}, 147 (1975).
  
\bibitem{Isgur:1978xi}
  N.~Isgur and G.~Karl,
  Phys.\ Lett.\  B {\bf 74}, 353 (1978).

\bibitem{Isgur:1978xj}
  N.~Isgur and G.~Karl,
  Phys.\ Rev.\  D {\bf 18}, 4187 (1978).

\bibitem{Isgur:1978wd}
  N.~Isgur and G.~Karl,
  Phys.\ Rev.\  D {\bf 19}, 2653 (1979);
  Erratum-ibid.\  D {\bf 23}, 817 (1981).
  

\bibitem{Glozman:1995fu}
  L.~Y.~Glozman and D.~O.~Riska,
  Phys.\ Rept.\  {\bf 268}, 263 (1996).

\bibitem{Glozman:1997ag}
  L.~Y.~Glozman, W.~Plessas, K.~Varga and R.~F.~Wagenbrunn,
  Phys.\ Rev.\  D {\bf 58}, 094030 (1998).

\bibitem{Glozman:1999vd}
  L.~Y.~Glozman,
  Nucl.\ Phys.\  A {\bf 663}, 103 (2000).

\bibitem{Bijker}
  R.~Bijker, F.~Iachello and A.~Leviatan,
  Annals Phys.\  {\bf 236}, 69 (1994);
  R.~Bijker, F.~Iachello and A.~Leviatan,
  Phys.\ Rev.\  D {\bf 55}, 2862 (1997).

\bibitem{Giannini}
  M.~M.~Giannini, E.~Santopinto and A.~Vassallo,
  Prog.\ Part.\ Nucl.\ Phys.\  {\bf 50}, 263 (2003);
  M.~Gorchtein, D.~Drechsel, M.~M.~Giannini, E.~Santopinto and L.~Tiator,
  Phys.\ Rev.\  C {\bf 70}, 055202 (2004).
  
\bibitem{Bonn} 
%
  U.~Loring, K.~Kretzschmar, B.~C.~Metsch and H.~R.~Petry,
  Eur.\ Phys.\ J.\  A {\bf 10}, 309 (2001);
%
  U.~Loring, B.~C.~Metsch and H.~R.~Petry,
  Eur.\ Phys.\ J.\  A {\bf 10}, 395 (2001);
  U.~Loring, B.~C.~Metsch and H.~R.~Petry,
  Eur.\ Phys.\ J.\  A {\bf 10}, 447 (2001);
%
  B.~Metsch,
  Eur.\ Phys.\ J.\  A {\bf 35}, 275 (2008).

\bibitem{Capstick:1992th}
  S.~Capstick and W.~Roberts,
  Phys.\ Rev.\  D {\bf 47}, 1994 (1993).

\bibitem{Roberts:2007ji}
  C.~D.~Roberts,
  Prog.\ Part.\ Nucl.\ Phys.\  {\bf 61}, 50 (2008).

\bibitem{Alexandrou:2009xk}
  {\it see e.g.} C.~Alexandrou,
  arXiv:0906.4137 [hep-lat], and references therein.

\bibitem{Isgur:1999jv}
  N.~Isgur,
  Phys.\ Rev.\  D {\bf 62}, 054026 (2000).


\bibitem{Chizma:2002qi}
  J.~Chizma and G.~Karl,
  Phys.\ Rev.\  D {\bf 68}, 054007 (2003).

%
\bibitem{Chizma:2004wr}
  J.~Chizma,
  Ph.D Thesis, The University of Guelph (2004), unpublished.

\bibitem{Capstick:2004tb}
  S.~Capstick and W.~Roberts,
  Fizika B {\bf 13}, 271 (2004).


\bibitem{Pirjol:2003ye}
  D.~Pirjol and C.~Schat,
  Phys.\ Rev.\  D {\bf 67}, 096009 (2003).
  
\bibitem{Nc}
  N.~N.~Scoccola, J.~L.~Goity and N.~Matagne,
  Phys.\ Lett.\  B {\bf 663}, 222 (2008);
  D.~Pirjol and C.~Schat,
  Phys.\ Rev.\  D {\bf 78}, 034026 (2008);
and references therein.

\bibitem{Pirjol:2008gd}
  D.~Pirjol and C.~Schat,
  Phys. Rev. Lett. {\bf 102}, 152002 (2009).

  
\bibitem{Old} 
        S. A. Dytman {\it et al.}, 
        Phys. Rev. C {\bf 51}, 2710 (1995);
%
        J. W. Price {\it et al.}, 
        Phys. Rev. C {\bf 51}, R2283 (1995);
%
        B. Krusche {\it et al.}, 
        Phys. Rev. Lett. {\bf 74}, 3736 (1995);
%
	A. Bock {\it et al.}, 
	Phys. Rev. Lett. {\bf 81}, 534 (1998);
%
	J. Ajaka {\it et al.},
	Phys. Rev. Lett. {\bf 81}, 1797 (1998); 
%
  F.~Renard {\it et al.},
  Phys.\ Lett.\  B {\bf 528}, 215 (2002).

\bibitem{Capstick:2007tv}
  S.~Capstick {\it et al.},
  Eur.\ Phys.\ J.\  A {\bf 35}, 253 (2008).

\end{thebibliography}
\end{document}